\documentclass{article}

\usepackage{amssymb}
\usepackage{amsmath}
\usepackage[dvips]{graphicx}

\begin{document}

\noindent 

\noindent 

\noindent 

\noindent 

\noindent \textbf{The Sky is for Everyone -- Outreach and Educaction with the Virtual Observatory}\\

\noindent 

\noindent \textit{Florian Freistetter (Astronomisches Recheninstitut, Universit\"at Heidelberg,}\\florian@ari.uni-heidelberg.de\textit{)}

\noindent \textit{Giulia Iafrate (INAF Osservatorio Astronomico di Trieste,}\\iafrate@oats.inaf.it\textit{)}

\noindent \textit{Massimo Ramella (INAF Osservatorio Astronomico di Trieste,}\\ramella@oats.inaf.it\textit{)}

\noindent \textit{and the AIDA-WP5 Team}\\\\

\noindent 

\noindent \textit{Summary: }The Virtual Observatory (VO) is an international attempt to collect astronomical data (images, simulation, mission-logs, etc), organize it and develop tools that let astronomers access this huge amount of information. The VO not only simplifies the work of professional astronomers, it is also a valuable tool for education and public outreach. For teachers and astronomers who actively promote astronomy to the public the VO is an great opportunity to access real astronomical data, use them and have a taste of the workaday life of astronomers.\\

\noindent 

\noindent \textit{Keywords:} Virtual Observatory, Education, Database, Public Outreach

\noindent\\ 

\noindent \textbf{Introduction}\\

\noindent 

\noindent Astronomy is a very attractive science for public outreach, teachers and students. It allows to practice very formative experiments/observations with relatively simple and inexpensive tools. Of course to have access to a telescope dramatically increases the interest of astronomy for the public and  schools. However, even if Internet has made available many resources on-line, public access to remotely controlled telescopes (e.g  Faulkes telescope and others) remain limited, if anything because time slots are limited and not easily scheduled during classroom hours.\\

\noindent 

\noindent In this paper we present the \textit{Virtual Observatory for Schools and Public} (the result of the Work Package 5 of the \textit{Astronomical Infrastructure for Data Access} project - AIDA-WP5). AIDA-WP5 is a free resource developed within the (European) Virtual Observatory project we describe briefly below . The aim of AIDA-WP5 is to give access to VO data using professional grade software tools modified to make them appealing and easily usable by students, teachers and public while maintaining the look and feel of  tools used by professional astronomers.\\

\noindent 

\noindent AIDA-WP5 is not simply a door to VO resources to be opened with VO tools: it is a self-consistent resource offering a set of use cases that are interesting astronomical problems to be solved using free software tools and data. Use cases are documents that include both a presentation of the astronomical problem and a guide to the use of VO tools and data. AIDA-WP5 can complement or even substitute access to real telescopes with obvious advantages of flexibility.\\

\noindent 

\noindent In the following sections we briefly describe the Virtual Observatory and, in particular the EuroVO-AIDA project; we describe in detail tools and use cases we developed for the outreach and education work package of the AIDA project. Finally, in the last section we give a short account of our direct experiences of using AIDA-WP5 tools in schools.\\

\noindent 

\noindent \textbf{The formation of a virtual observatory}\\

\noindent 

\noindent In ancient times, astronomers where looking at the sky with their naked eyes and noted their observations on clay tablets, parchment, papyrus and paper. When Galileo Galilei introduced the telescope to astronomy, this did not change: astronomical observations and scientific results were published and stored in books and papers. When photographic plates came into common use, observatories had to store them too because they constituted the raw data and formed a valuable scientific archive. Nowadays we image the sky directly to a file on a computer and store our data digitally. Observatories all over the world, together with astronomical satellites, probes and telescopes in space,  produce vast amounts of electronic data every day and night.\\

\noindent 

\noindent In the past, accessing the collection of photographic plates of a certain observatory was difficult. One had to travel to the place itself to inspect the images or had to ship the plates which took a long time exposing plates to possible damage or destruction. Today instead, exchanging digital data is very easy. Internet allows an unproblematic and fast interchange of scientific data.\\

\noindent 

\noindent Thus, thanks to the Internet, every astronomer could, in principle, easily access and profit from the observations made by all other astronomers worldwide. In practice however, a complex infrastructure is needed to collect and distribute the multitude of astronomical data. Since data are stored with different formats and according to different standards, internet communications and exchanges have to obey protocols of communication and pass several processes of verification. This infrastructure is provided by the Virtual Observatory (VO).\\

\noindent 

\noindent The International Virtual Observatory Alliance (IVOA) was established in 2002. IVOA comprises now 17 VO projects from Armenia, Australia, Brazil, Canada, China, Europe, France, Germany, Hungary, India, Italy, Japan, Korea, Russia, Spain, the United Kingdom, and the United States. Its mission is to\\

\noindent 

\noindent \textit{"facilitate the international coordination and collaboration necessary for the development and deployment of the tools, systems and organizational structures necessary to enable the international utilization of astronomical archives as an integrated and interoperating virtual observatory."\footnote{ For details see http://ivoa.net/}}\\

\noindent 

\noindent In order to explain some of the reason for setting up a VO, consider the following. Often, if not always, images taken by one observer could be also of great value for another astronomer who is researching a totally different topic. Normally, the two astronomers would probably never be aware one of the other and the second scientist would perform his own observations thus producing a second set of the same data already archived by the first. Thanks to the VO, the images of the first observer can be easily found and used by the second astronomer with a significant increase of efficiency and reduction of costs.\\

\noindent 

\noindent Ultimately, the goal of the VO project is to provide a skin under which to hide the complexities of a variety of data coming from different instruments, different telescopes, different data centers: as seen by an astronomer, the VO should look like a normal telescope. In fact, astronomers may not even know whether they are using a real or virtual instrument.\\

\noindent 

\noindent Scientists are not the only group that should profit from the Virtual Observatory. Amateur Astronomers can access  professional data via the VO and use it for their work. Of course they are able to submit their own observations, thus contributing directly to scientific progress. \\

\noindent 

\noindent Most importantly, the VO is a great opportunity for teachers, students and in general for the public. Most data in the VO is available to \textit{all} people -- astronomers or not - and in principle everybody should be able to access the same scientific data and tools as professional astronomers. However, without proper explanations, professional data and specialized tools are of little use for laypersons and non professional astronomers.  The VO offers  a great opportunity  to astronomers who actively promote their science to introduce people to real astronomical data, letting them experience a little part of the everyday life of astronomers.\\

\noindent 

\noindent \textbf{Euro-VO AIDA for Outreach and Education}\\

\noindent 

\noindent In the framework of the European Euro-VO AIDA (Astronomical Infrastructure for Data Access) project\footnote{ For details on EURO-VO AIDA see http://www.euro-vo.org/}, which is funded by the European Commission under the Research Infrastructure FP7, a special effort is made in this direction. The 5th of AIDA's seven work packages is dedicated to develop tools and methods to let students, teachers and in general the public benefit from the European investment in the VO.

\noindent 

\noindent In a first step, we chose existing professional software tools for the retrieval, visualization and analysis of VO data in order to adapt them for educational and outreach purposes. One of the most popular tools to access the VO is the program "Aladin", developed by the Centre de données astronomiques de Strasbourg (CDS). In its professional version Aladin is too complicated and contains too many specialized functions to be of any interest for non-professional users. As a consequence, we transformed Aladin to a more comprehensible version.

\noindent 

\begin{figure}[]
\begin{center}
\includegraphics[width=4.5in]{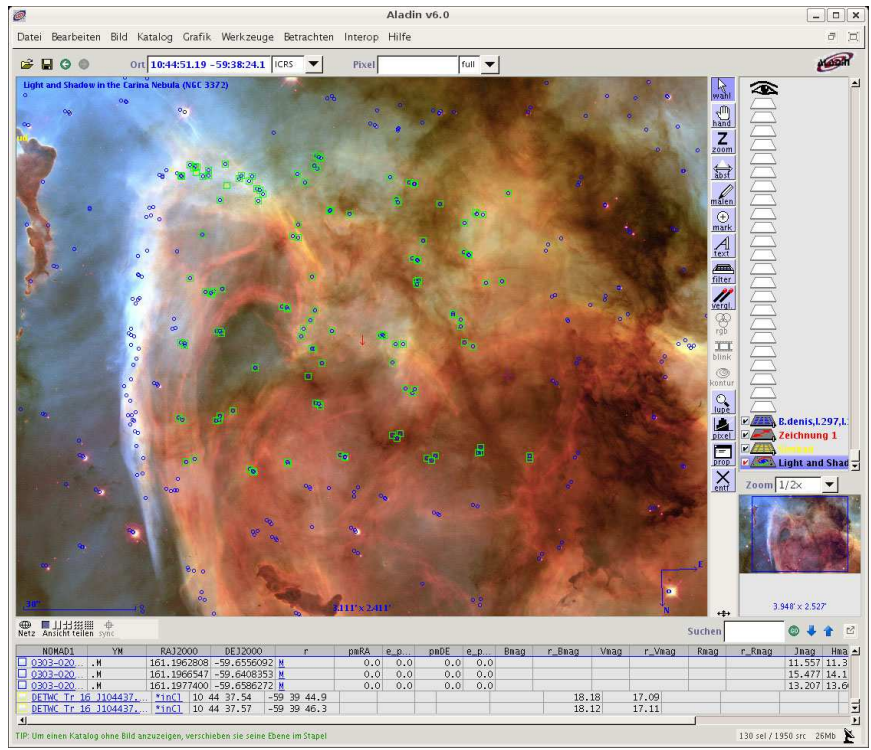}
\end{center}
\caption{The VO-software Aladin; depicting an image of the Carina Nebula taken by the Hubble Telescope with an overlay of catalogue data from the Vizier Database (Credit: Authors, NASA, ESA, CDS)}
\label{fig1}
\end{figure}

\noindent

\noindent A second valuable tool used and modified by AIDA is the sky browser Stellarium, developed by the European Southern Observatory (ESO). It allows people to depict the sky at any give location and for an given date and watch the motion of the stars and planets.

\noindent 

\begin{figure}[]
\begin{center}
\includegraphics[width=4.5in]{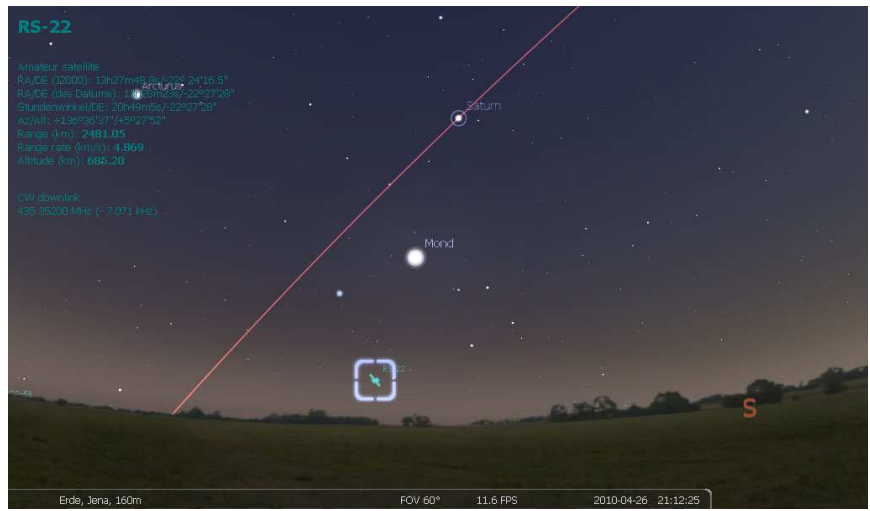}
\end{center}
\caption{Stellarium shows how the sky will look like. The moon, a satellite and the orbit of Saturn is visible  (Credit: Authors, ESO)}
\label{fig2}
\end{figure}

\noindent 

\noindent Using and adapting Aladin and Stellarium it was our goal to develop tools that enable everyone - and not only professional astronomers -  to (virtually) observe the sky and access all relevant data. For this purpose, it was not only necessary to provide the software; there was also a need for examples and use-cases that demonstrate the usage of Aladin and Stellarium in an easy and comprehensible way. Thus besides the development of the software it was our task to collect and create such examples that show how the data in the VO can be accessed and used.\\

\noindent 

\noindent We chose the use-cases in order to  apply them in schools, universities and public outreach. The VO is a great opportunity for teachers to introduce the students to real astronomical data and the methods to work with it. We have developed a series of such use cases of different complexities that are adequate for students of different ages and deal with different astronomical topics ranging from the distribution of asteroids to the distance of the galaxies.\\

\noindent 

\noindent A typical use-case that can be employed in a school or an beginners astronomy lecture at a university is dealing with a concrete topic, like the determination of the distance of the Andromeda Galaxy. Every use case starts with a general introduction that, in this example, explains how astronomers are able to measure distances and that not even 100 years ago, we did not know if our Milky Way was all there is in the universe or if the faint "nebulas" one was observing in the sky could be distant islands of stars as our own galaxy. To resolve that dispute, on had to measure their correct distance. This was done by Edwin Hubble in 1924 by utilizing a special relation between the brightness and the period of certain variable stars known as Cepheids. The discovery of Hubble that the Andromeda "Nebula" was in fact an extremely distant galaxy full of stars and that our universe consisted of myriads of such galaxies that all seem to move away from us the faster the further away they are was revolutionary and changed astronomy and the way we view the world. Making use of the tools and data from the VO, it is easy to retrace the steps of Hubble using real astronomical data.\\

\noindent 

\noindent In our use case we show that with Aladin, one can not only access many astronomical images but also a vast number of stellar measurements and catalogs. It is easy to retrieve observational data of all Cepheid stars in the Andromeda Nebula and use the built-in spreadsheet tools of Aladin to process that measurements in the same way as Hubble did when he was calculating the distance of Andromeda.\\

\noindent

\begin{figure}[]
\begin{center}
\includegraphics[width=4.5in]{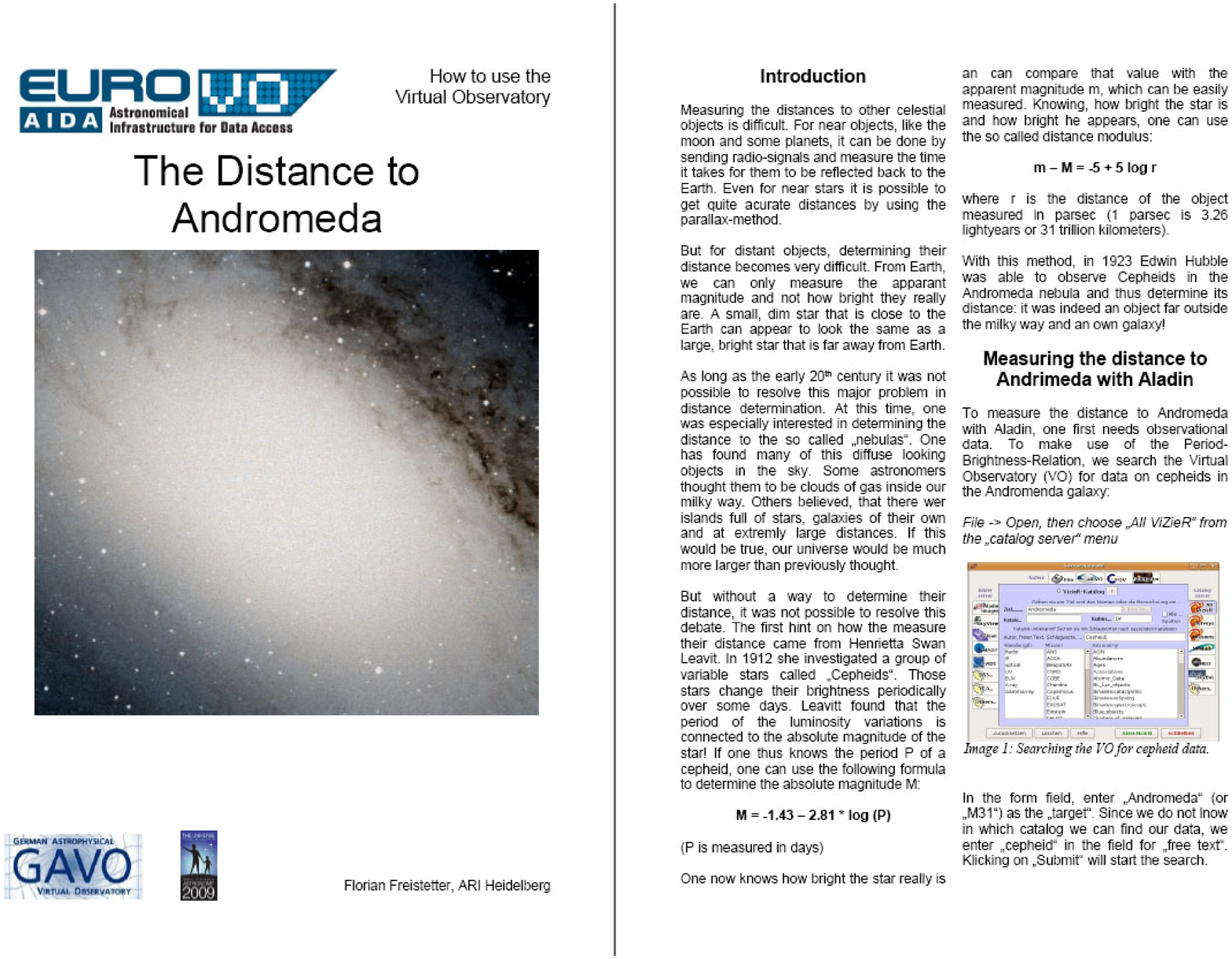}
\end{center}
\caption{Example of an use case: measuring the distance to Andromeda with Aladin (Credit: Authors}
\label{fig3}
\end{figure}

\noindent 

\noindent Other Aladin use-cases developed by the AIDA-team deal with the motion of stars, the confirmation of supernovae or the properties of stars in the Pleiades\footnote{ All use-cases and all software can be downloaded from http://wwwas.oats.inaf.it/aidawp5}. For younger students or the general public who are not able to or not interested in doing astronomical calculations, we have developed other use cases that make use of the Stellarium sky browser.\\

\noindent 

\noindent One such example deals with the wide spread myth of a world ending catastrophe on December 21st 2012; popularized all over the world as main theme of Roland Emmerichs blockbuster-movie "2012". A major claim of the 2012-doomsayers is, that exactly on 12/21/2012 the planets of the solar system will align perfectly to form a straight line and the resulting gravitational perturbations will disrupt the Earth or at least cause major catastrophes (floods, earthquakes, etc). This claim can easily be refuted by using Stellarium.\\

\noindent 

\noindent In our use case we again start with a general introduction that explains how the planets in the solar system move and how this results in the astronomical phenomenon of conjunctions. We then show how one can depict the position of the planets for any given time and place and give examples of interesting conjunctions in the past (e.g. the conjunction of May 2000 or the conjunction of Jupiter and Saturn in the year 7 BCE that may be the base of the story of the "Star of Bethlehem"). We also show that it is easy to confirm that there will be no special alignment in the year 2012 and give instructions on how to calculate the (negligible) gravitational effect on Earth if there ever would be such a conjunction.\\

\noindent 

\begin{figure}[]
\begin{center}
\includegraphics[width=4.5in]{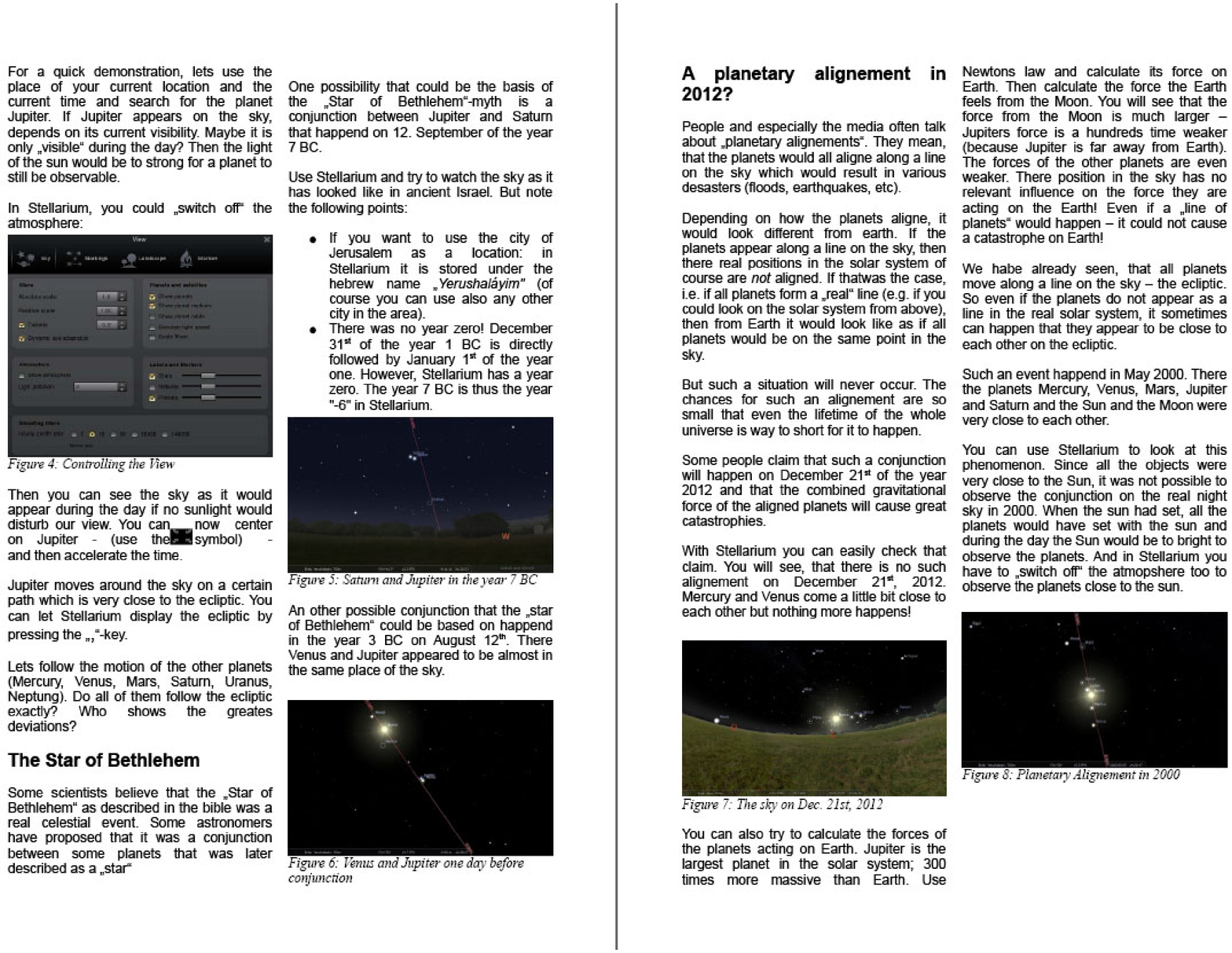}
\end{center}
\caption{Example of an use case: learning about planetary conjunctions in Stellarium (Credit: Authors)}
\label{fig4}
\end{figure}

\noindent 

\noindent \textbf{Classroom Experiences with AIDA-WP5}\\

\noindent 

\noindent Led by the Astronomical Observatory of Trieste, the AIDA use cases were applied and tested in many Italian schools with students aged 14 and 18 -- with great success! Four hours of teaching were dedicated to each use case: one hour each to introduce the astronomical background and the concept of the VO and two hours were reserved for the students to actually work on the problems\footnote{ For details see Iafrate G., Ramella M., Boch. T., Bonnarel F., Chéreau F., Fernique P., Freistetter F.: "Un progetto didattico per le scuole secondarie : EuroVO-Aida/WP5``, Giornale di astronomia. 2010/1. 31 or Ramella M., Iafrate G., Boch T., Bonnarel F.,Chéreau F., Fernique P., Freistetter F.: "At School with the European Virtual Observatory`` in "Proceedings of the IAU Symposium Astronomy and its Instruments-Before and After Galileo``, 2010} . We collected over 250 feedback-forms that were very valuable in order to advance our tools and use-cases. Figure 5 shows some results of the evaluation.

\noindent 

\begin{figure}[]
\begin{center}
\includegraphics[width=4.5in]{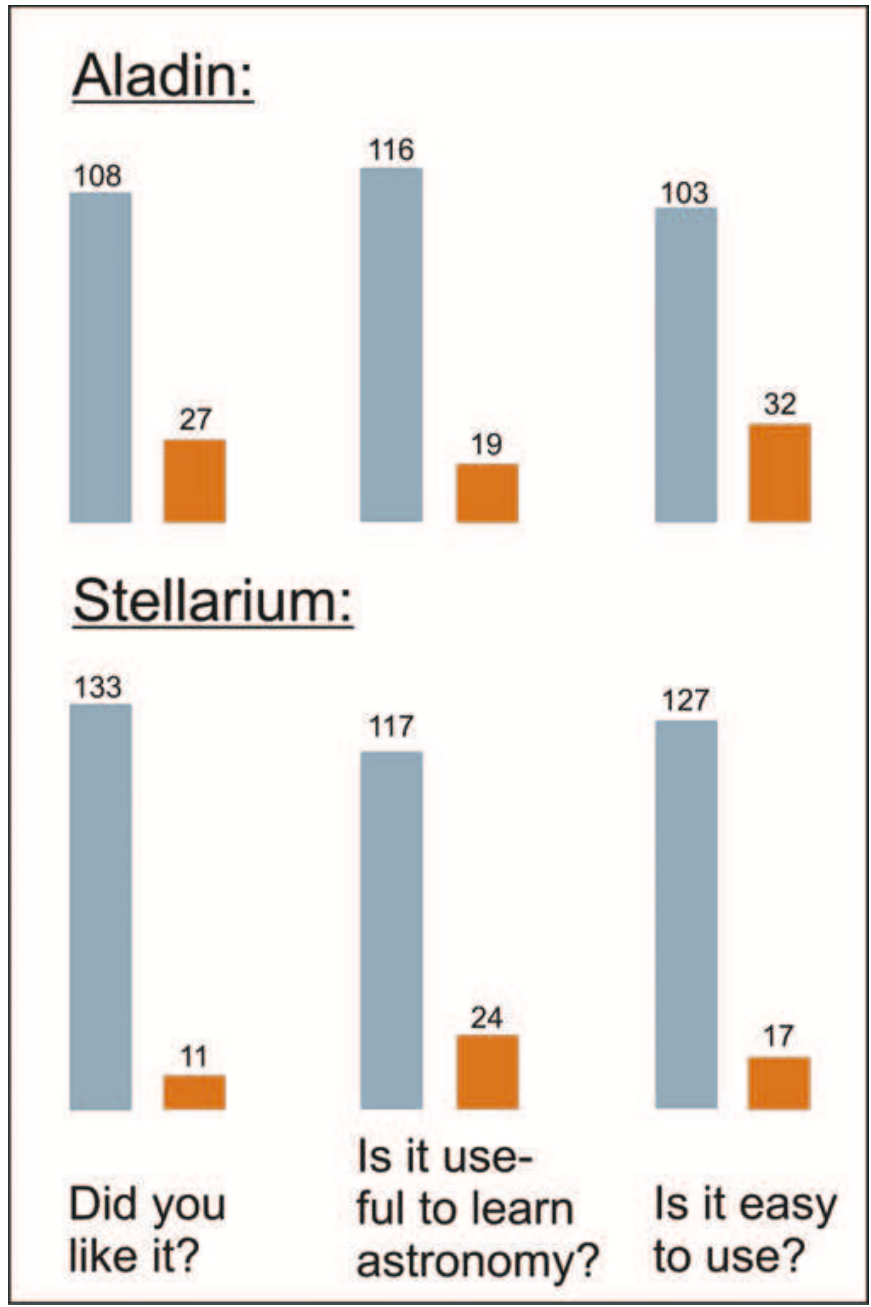}
\end{center}
\caption{Evaluation of Aladin and Stellarium. Blue bars give the number of students who answered with "Yes"; orange is the number of people that answered with "No" (Credit: Authors)}
\label{fig5}
\end{figure}

\noindent 

\noindent Additional input came from various groups of amateur astronomers. Currently, a second campaign is running in order to see if teachers can work with the use cases without help and supervision by professional astronomers.\\

\noindent 

\begin{figure}[]
\begin{center}
\includegraphics[width=4.5in]{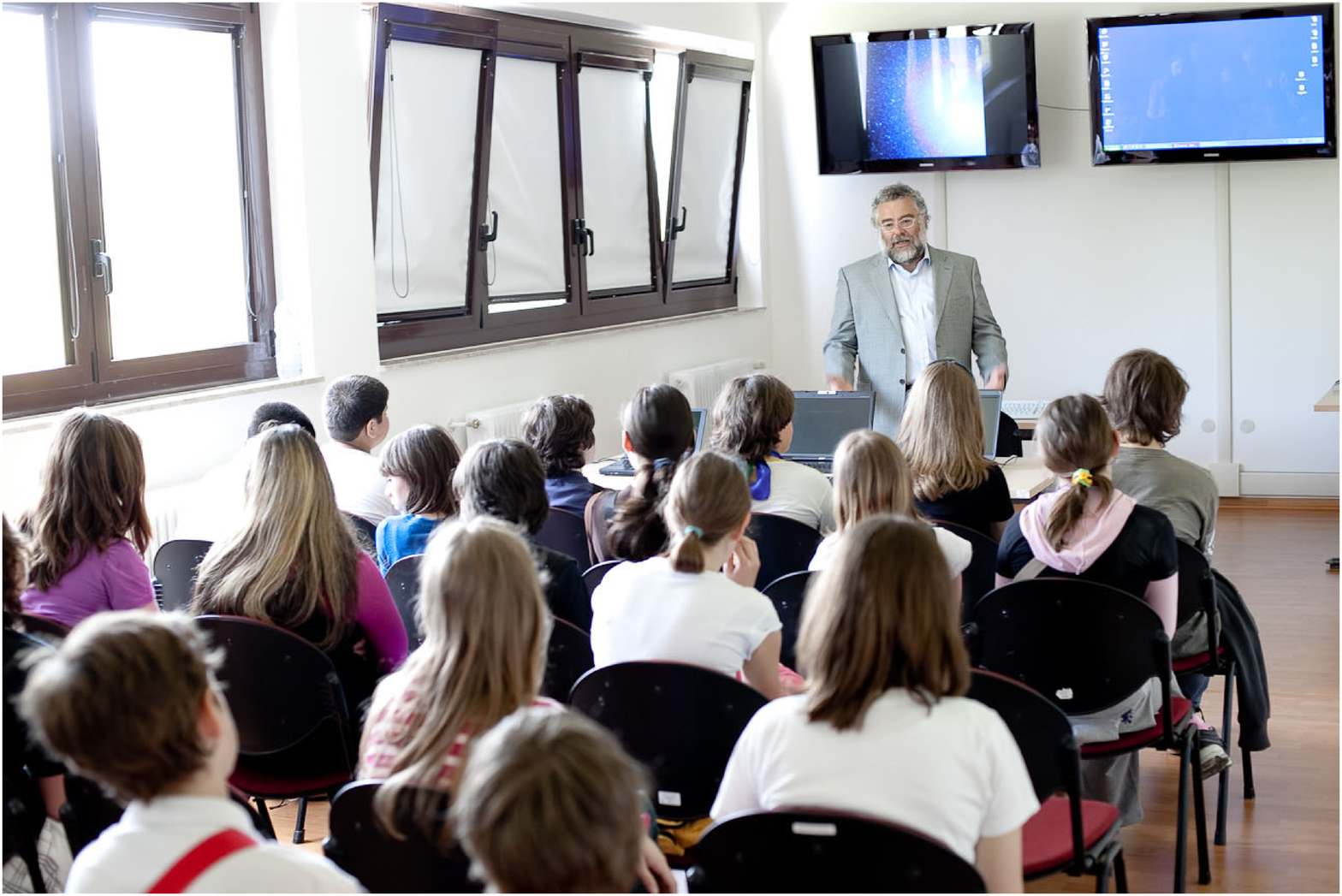}
\end{center}
\caption{Massimo Ramella (OATS) is introducing the Virtual Observatory to students of an italian school (Credit: Authors)}
\label{fig6}
\end{figure}

\begin{figure}[]
\begin{center}
\includegraphics[width=4.5in]{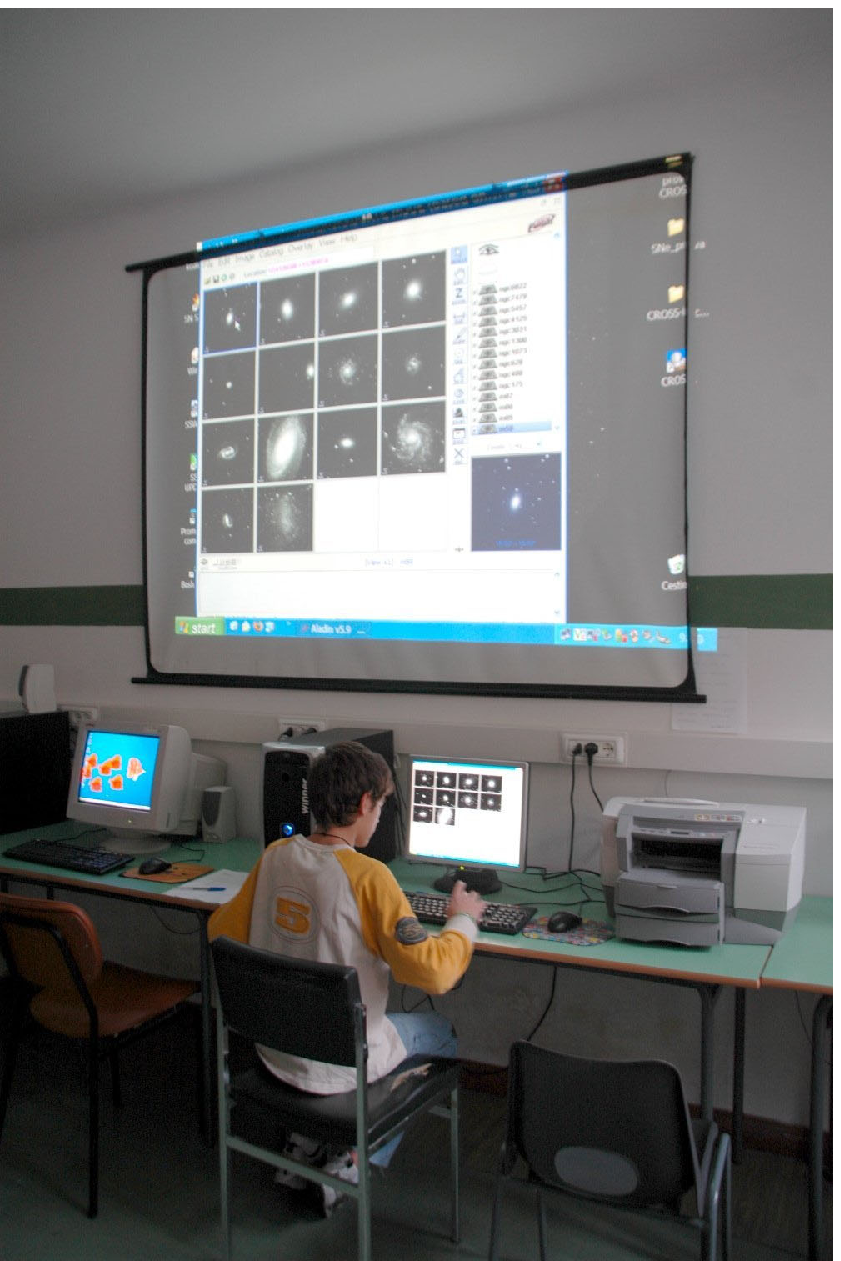}
\end{center}
\caption{A student is working with the Virtual Observatory (Credit: Authors)}
\label{fig7}
\end{figure}

\noindent \textbf{Conclusions}\\

\noindent 

\noindent Astronomy is a science that fascinates not only professional astronomers but also the general public. The AIDA-WP5 project is an attempt to make the large collection of astronomical data accessible and understandable for everyone who is interested in the sky. The goal is to obtain a set of dedicated tools and examples that can be used by teachers at schools and universities, by amateur astronomers and people working in public outreach in order to understand the concept of the VO and deploy it autonomously. The Virtual Observatory should become a standard tool not only for professional astronomers, simplifying their work, but also for everybody  who wants to introduce people to the vast amount of knowledge and beauty that is uncovered by astronomical research.\\

\noindent \textit{}

\noindent 

\noindent \textbf{Biography:}\\

\noindent \textbf{}

\noindent \textit{Florian Freistetter} is an astronomer, working for the European Virtual Observatory EURO-VO at the \textit{Astronomisches Recheninstitut} of the University Heidelberg (Germany). Before, he has investigated the dynamics of asteroids and extrasolar planets at the observatories of the universities of Vienna and Jena. He is the author of the ScienceBlog \textit{Astrodicticum Simplex}\\ (http://www.scienceblogs.de/astrodicticum-simplex/)\\

\noindent 

\noindent \textit{Giulia Iafrate} works on astronomy outreach and education at the Astronomical Observatory of Trieste (Italy). She cooperates also with the Italian National Institute for Nuclear Physics in the analysis of the data of the Fermi-LAT satellite.\\

\noindent 

\noindent \textit{Massimo Ramella} is associate astronomer at the INAF-Osservatorio Astronomico di Trieste (OATs). He coordinates the outreach and education activities of the OATs. He is the team leader of work package 5 of the Euro-VO AIDA project. His field of research includes  the large scale structure of the universe and systems of galaxies.

\noindent 

\noindent

\end{document}